\documentclass[allclo,superscriptaddress,eqsecnum,amsfonts,showpacs]{revtex4}
\usepackage{epsfig}

\newcommand{\be}{\begin{equation}}
\newcommand{\ee}{\end{equation}}
\newcommand{\bea}{\begin{eqnarray}}
\newcommand{\eea}{\end{eqnarray}}

\newcommand{\ep}{i\varepsilon}
\newcommand{\nn}{\nonumber}

\newcommand{\kv}{|\vec{k}|}
\newcommand{\pv}{|\vec{p}|}

\begin{document}

\title{General method of solution of Schwinger-Dyson equations in Minkowski space}

\author{Vladimir ~\v{S}auli}
\affiliation{CFTP 
IST Lisbon Portugal, and OTF UJF Rez near Prague, CZ}

\author{Zoltan Batiz}
\affiliation{CFTP 
IST Lisbon, Portugal}

\begin{abstract}
Novel solutions of Minkowski QED2+1 and QCD Schwinger-Dyson equations will be presented and
discussed. The resultant propagators of confined degrees of freedom will be shown.
\end{abstract}

\maketitle
 
\section{Motivation and Introduction}

In perturbation theory  these are always the free propagators which enter into the Feynman diagrams 
hence the well known calculation trick known as the  Wick rotation is always feasible. 
In this way the Euclidean and real world -Lorentz (Minkowski)- space results are related
through  a well defined analytical continuation. 
As a consequence of the perturbative technique, looking in momentum space, the Green`s functions (GFs) of given quantum field theory have a real branch points whose positions can be traced by analyzing the  matching propagator poles in their possible convolutions inside the Feynman integrals. The location of such branch points
corresponds with physics -- with  opening particle threshold production. For the momenta above threshold the GFs become complex with real and imaginary parts  uniquely  related by well known dispersion  and unitarity relations. Assuming  particles can freely propagate, the appropriate propagators must have a real pole and the unitarity relations remain to be valid for  S-matrix elements. This, of course completely follows the idea of LSZ reduction formula which relates S-matrix and Green`s functions of given theory.

Apart of genuine  perturbation theory success in quantum electrodynamics and electroweak sector of the Standard Model,  hadrons built of "light" quarks gain their masses through the approximate dynamical chiral symmetry breaking (DCSB) while  there is simultaneously a  nonperturbative mechanism which  forbids colorless hadrons  to  disintegrate to any free quarks (or colored states in general).
Similarly, the brehmstrahlung and consequential propagation of an on-shell gluon could be similarly impossible  as well.  Clearly the phenomena of DCSB and color confinement lie beyond the access of the perturbation theory, which is based on particle-field duality and which starts to work with free propagators from the very beginning. 

On the other hand the GFs of confined fields are basic element to build hadronic wave functions
form the first principle, i.e. using the original QCD Lagrangian degrees of freedom- the quark and gluon fields.
The spectra of all hadrons have timelike signature $P^2>0$ and the masses and and decays are identified from  measurements performed in the real world- one time and three space coordinates forming the Minkowski space. The intentions of my talk is to mention various non-perturbative methods which can be or have been used to evaluate GFs in strong coupling theories in  Minkowski space. The natural framework is based on the setup of QCD Schwinger-Dyson equations
\cite{ROBWIL1994,ALKSME2001,BINPAP2009}.

The standard philosophy of nonperturbatively solved SDEs is common with the lattice theory.
This is to start with Euclidean space definitions of Greens functions and analytically continue the Euclidean results to the Minkowski space. This is clearly always possible, even if the observed singularities
do not reflect usual assumptions imposed by Wick rotation \cite{ESTRADA}. The SDEs are selfconsistent- what enters SDEs is output as well- and the selfconsistency of SDEs is more important in QCD then perturbative analyticity, which is not 
guaranteed in a confining theory at all. In other words, what can be obtained by analytical continuation of Euclidean 
GFs may not agree with the direct Minkowski solution. That the other then usually expected singularities appear when making continuation from Euclidean towards Minkowski space  is a well known fact \cite{MAR1,MAHO1992}. However, in order to be able to correctly  check the assumptions, one should be able to compare the results  based on the continuation  of the Euclidean models with the results of the same (as possible) model directly solved in the Minkowski space. 
In this talk I present a direct solution of the SDEs in Minkowski space, the first one for QED2+1 , where no analytical assumptions are made. The second model I present here is large $N_f$ QCD where some assumptions are necessary in order to make a Minkowski solution  possible.

Before doing so I mention some Minkowski SDEs solution which are based on the assumption of validity of Wick rotation.

\section{Semi-perturbative methods based on integral representation}

Assuming perturbative analyticity of Green`s function one can impose
Khallen-Lehman representation for propagators
\be \label{jednac}
G^{KLR}(k)=\int\limits_0^{\infty}d M^2 \frac{\rho(M^2)}{k^2-M^2+\ep} 
\ee
with the important property
\be
\int\limits_{-\infty}^{\infty} dk^2 G^{free}(k)=-i\pi .
\ee

Quite independently on the details of the models the momentum SDEs can be turned  into a "regular"
equations for the continuous part of the  Lehmann weight $\rho_c(M^2)$ (assuming that the full $\rho$ includes also  a single delta function corresponding with mass pole of the propagator).
Using the Feynman tricks and some standard algebra one can  always arrive to the dispersion relation  for the inverse propagator:
\be \label{dvac}
G^{-1}=polyn. +\int\limits_T^{\infty}d M^2 \frac{\sigma(M^2)}{k^2-M^2+i\ep} \, ,
\ee
where the selfenergy (mass,polarization, etc.) weight function nontrivially depends on $\rho$.

Propagators are complex above the threshold  T and the both $\rho$ and $\sigma$ can be extracted by comparing Re and Im parts of G (for a review see \cite{SAULI}), the other application of the spectral technique has found its place in the pinch technique study of Yang-Mills theories \cite{cornwall2009,AGBIPA2008,PAPA}.

For such Minkowski space calculation  the results were always
available for small coupling wherein they are quantitatively comparable with the perturbation theory  (note, by  construction the perturbative contribution is always involved in the game), however for a large  enough coupling there is observed some disagreement with the assumption  \cite{SAUBI}, or even the solution is not feasible at all 
\cite{SAULI}. Reconstruction of GFs in the  spacelike domain, where $p^2<0$ in our metric (recall $G,\rho, \sigma$ are primarily obtained  at timelike region ($p^2>T$), see \cite{SAULI}), then  using independently $\rho$ in (\ref{jednac}) and $\sigma$ in (\ref{dvac}) serves as a simple test of validity of the assumptions. Regarding modeling of QCD GFs the observed discrepancy is mild for pinch technique gluon propagator presented in  the paper \cite{SAULI2009} but becomes  a disaster when attempting to find a spectral solution for the quark propagator  \cite{SAUBI}.   These observations suggest that perturbative analyticity is too strong assumption for the QCD quark propagator and very likely  for gluon propagator as well (to be strict, the position of unusual singularities necessarily affects the analytical structure of the other GFs, since they are necessarily related through the SDEs).

\section{QED2+1}

QED2+1 is known \cite{APP1986},\cite{APP1988} to be a simple confining theory which for a small number of flavors posses DCSB. For static (heavy) electrons it posses a logarithmic confinement if the unquenching effect is not going to spoil the single pole of photon propagator \cite{FADM2004qed3,BRCR2008}. QED2+1 is also  pedagogical example of numerically soluble theory in Minkowski space in  ladder approximation  of electron SDE \cite{SAUBAT2009}. Here we employ Landau gauge and 
 $A=1$ approximation (which is exact in Euclidean space, but not obvious or clear in Minkowski space).  
In addition what I presented in the talk  I provide  a simple derivation  for completeness here.

The full electron propagator reads
\bea
S=S_s(p^2)\not p&+&S_v(p^2) 
\\ 
S_s=\frac{B(p)}{A^2(p)p^2-B^2(p)}\, \, &;& 
S_v=\frac{A(p)}{A^2(p)p^2-B^2(p)} \, \, , \nn 
\eea
where in our approximation
\bea
B&=&m+\Sigma_B\,\,\, , \, \,  A=1 \, \, ;
 \\
\Sigma_B&=&-2ie^2 \int\frac{d^3k}{(2\pi)^3} S_s(k^2)G(q^2)\, \, ,
\label{oursde} 
\eea
where $q=k-p\,\, ; G=1/q^2$. We intend to rewrite Eq. (\ref{oursde}) into a numerically easily soluble integral equation. In order to avoid numerical interpolation we shall use the arguments of propagators as  integral variables. Here we perform the derivation for the timelike 
 region of $p^2$, and without  loss of generality we can use the frame where $p^{\mu}=(p,0,0)$.
For this purpose we consider a bit more general loop integral.
\be \label{secret}
I[S,G,n;p^2)\equiv i\int\frac{d^3k}{(2\pi)^3} S(k^2)G(q^2)(k.q)^N \, ,
\ee
where $q=p-k$ and where the functions $S,G$ are  functions only of variable $k^2$ and $q^2$ respectively and clearly $N=0$ is sufficient in  our case. It is obvious that all one loop Lorentz invariant selfenergy contributions can be written in terms of considered integral (in any theory). For the  space part of the Lorentz three-vector we can use the usual spherical coordinates, so the scalar product $k.p$ reads
\be
k.p=k_0p_0-\kv\pv cos\phi\, ,
\ee
and for the measure we have
\be
\int d^3k=\int_{-\infty}^{\infty}dk_0\int d\kv\kv\int_{0}^{2\pi}d\phi\, .
\ee
As we choose the argument  of $S$ as an integration variable, the integral (\ref{secret}) can be written in the following way
\bea \label{mezi}
&&\frac{i}{(2\pi)^2}\int_0^{\infty}d\kv\kv\int_{-\kv}^{\infty}\frac{dk^2}{2\sqrt{k^2+\vec{k}^2}}
\sum_{i=\pm}S(k^2)G(q^2_i)(k.q)^N \, \, ;
\\
&&q^2_\pm=k^2+p^2\mp 2p\sqrt{k^2+\kv^2} \, \, ;
\\ \nn
&&(k.q)_{\pm}=k^2\mp p\sqrt{k^2+\kv^2}\nn \, \, ;
\eea
where we have used the explicit form of  $p^{\mu}$. 
In the next step we split the integral over $k^2$ to its spacelike and timelike subregions
and change the order of  integration
\be
\int_0^{\infty}d\kv\kv\int_{-\kv^2}^{\infty} dk^2\rightarrow
\int_{-\infty}^{0}dk^2\int_{\sqrt{-k^2}}^{\infty} d\kv +\int_0^{\infty}dk^2\int_0^{\infty}d\kv \, \, .
\ee
Consequently we make substitution $\kv \rightarrow q^2_{\pm}$ in the first (for the "plus" index) and in the second (for the "minus" index) term in (\ref{mezi}) separately, thus getting for $I$  
the following expression:
\bea
\frac{(2\pi)^2}{i}I&=&\int_{-\infty}^{0}dk^2\int_{k^2+p^2}^{-\infty} dq^2_+
S(k^2)G(q_+^2)\frac{(1/2(q^2_++k^2-p^2))^N}{-4p}
\nn \\
&+&\int^{\infty}_{0}dk^2\int_{k^2+p^2-2\sqrt{p^2k^2}}^{-\infty} dq^2_+
S(k^2)G(q_+^2)\frac{(1/2(q^2_++k^2-p^2))^N}{-4p}
\nn \\
&+&\int_{-\infty}^{0}dk^2\int_{k^2+p^2}^{\infty} dq^2_-
S(k^2)G(q_-^2)\frac{(1/2(q^2_-+k^2-p^2))^N}{4p}
\nn \\
&+&\int^{\infty}_{0}dk^2\int_{k^2+p^2+2\sqrt{p^2k^2}}^{\infty} dq^2_-
S(k^2)G(q_-^2)\frac{(1/2(q^2_-+k^2-p^2))^N}{4p} \, .
\eea
In what follows we use a more compact notation by relabeling $q^2_-\rightarrow q^2$ and
 $q^2_+\rightarrow q^2$, further we can rewrite the boundaries in a fully equivalent manner 
 \bea
\frac{(2\pi)^2}{i}I&=&\int_{-\infty}^{\infty}dk^2\int_{-\infty}^{\infty} dq^2
S(k^2)G(q^2)\frac{(1/2(q^2+k^2-p^2))^N}{4p}
\nn \\
&-&\int^{\infty}_{0}dk^2\int_{k^2+p^2-2\sqrt{p^2k^2}}^{k^2+p^2+2\sqrt{p^2k^2}} dq^2
S(k^2)G(q^2)\frac{(1/2(q^2+k^2-p^2))^N}{-4p}
\eea
which is the most general expression whether  $G$ is known or not.

  $G$ is the free boson propagator in our approximation for which $<G>=0$, so the first term vanishes (we did not consider $\ep$ prescription here, as it has no important effect in this case)  and  the $q^2$ integration can be performed analytically leading to the final expression for  (\ref{oursde})
\be \label{result}
B(p^2)=m+\frac{ie^2}{4\pi^2}\int^{\infty}_{0} dk^2 S_s(k^2)ln\frac{|k-p|}{(k+p)}.
\ee
Recall that the same has been derived in \cite{SAUBAT2009} by using a hyperbolic coordinates.

We assume the propagator functions are complex almost everywhere (i.e. there are no perturbative thresholds)
\bea \label{cpxS}
S_s&=&\frac{B(p)}{A^2(p)p^2-B^2(p)}=\frac{R_B c_1+\Gamma_B c_2}{c_1^2+c_2^2}
+i\frac{\Gamma_B c_1-R_B c_2}{c_1^2+c_2^2}\, ;
 \\
S_v&=&\frac{R_A c_1+\Gamma_A c_2}{c_1^2+c_2^2};
+i\frac{\Gamma_A c_1-R_A c_2}{c_1^2+c_2^2}
\nn \\
c_1&=&(R_A^2-\Gamma_A^2)p^2-(R_B^2-\Gamma_B^2);
\nn \\
c_2&=&(R_A\Gamma_A p^2-R_B\Gamma_B^2) \nn;
\eea
where $R,\Gamma $ are Re an Im parts of proper GFs, $A=R_A+i\Gamma_A \, ; \, B=R_B+i\Gamma_B$.

In this way we get two coupled integral equation for $R_B,\Gamma_B$ which can be solved by the method of iterations.
These  integral equations are  regular in $S$ unless $\Gamma_B$ does not vanish.

\section{QED2+1 versus QED3}

At this point it is quite interesting to compare formally obtained expression for Minkowski  $B$
with its Euclidean counterpartner.
Stress the both variables $k,p$ in (\ref{result}) are  timelike fourmomenta since the timelike solution decouple from the spacelike one.  
Furthermore as we will see $S_s$ given by Eq. (\ref{cpxS}) is entirely complex function in the timelike Minkowski subspace.

The conventionally written  Euclidean partner (note $p_E^2=-p^2>0$ ) is a smooth regular function
\be
S_s(p_E^2)=\frac{B(p_E^2)}{A^2(p_E^2)p^2_E+B^2(p_E^2)}\nn\, .
\ee
When using the same approximation it satisfies Wick rotated Euclidean ladder SDE:
\be
B_E(p^2_E)=m-\frac{e^2}{4\pi^2}\int\limits_0^{\infty} dk^2_E S_s(p_E^2)
ln\frac{|k-p|}{(k+p)} 
\ee
where $B_E$ is numerically known. It is real, and it stays real even if one allows  for a complex valued propagator (imaginary part is not generated), B  is known to be nontrivial for any coupling $e$  
which includes classical chiral symmetry  ($m=0$) case.

Due to the very simple structure of SDEs we can find recipe how to get the first from the second one, however one cannot say that the first is the analytical continuation of the second.
Very formally, the Euclidean ladder SDE can be get by using the contour shown in Fig.1.
\begin{figure}
\centerline{\epsfig{figure=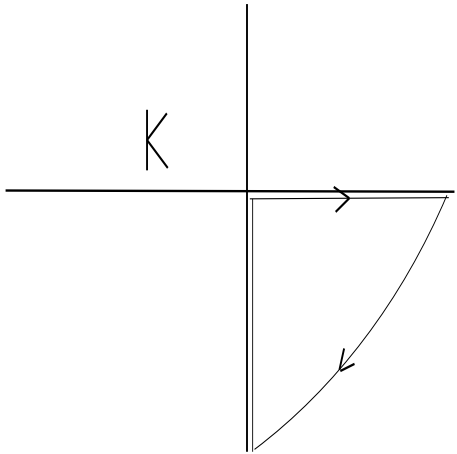,width=6truecm,height=6truecm,angle=0}}
\caption[right]{\label{contr}Graphical explanation of   Minkowski --> Euclidean continuation for ladder QED2+1. The figure shows the complex plane of  the integral variable $\sqrt{k^2}$ used in the 1-dim integral SDEs in Minkowski and Euclidean spaces. The recipe that gives the correct continuation is based  on the integration along the drawn contour such that we take $\int_{Im k} \rightarrow \int_{Re k} $ we going from the first to the second space, albeit Cauchy lemma for such contour cannot be used, since the function is not holomorphic in the interior of the shown curve. The arc does not contribute as radius goes to infinity because of Jordan lemma.}
\end{figure}
with simultaneously continuing external variable $p$ to the spacelike region as well (explicitly $p_E\rightarrow -ip_M $.

\begin{figure} 
\centerline{\epsfig{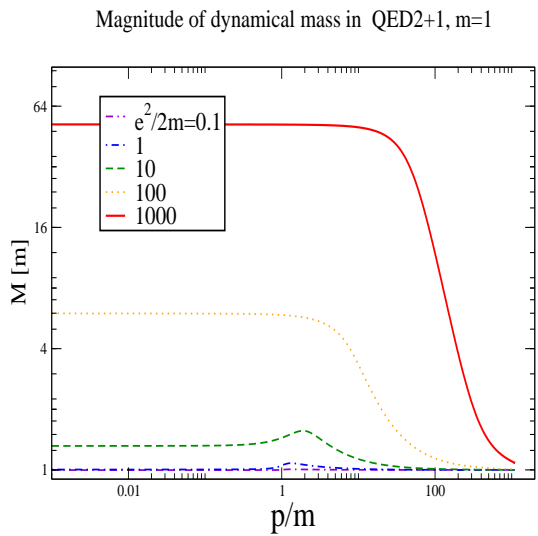}}
\caption[caption]{\label{jedna} Magnitude of electron dynamical mass function}
\end{figure}

\begin{figure}
\centerline{\epsfig{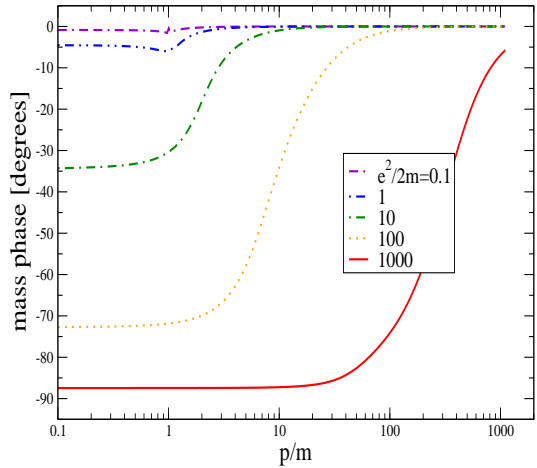}}
\caption[caption]{ \label{dva} Phase of dynamical electron mass $M=|M|e^{i\phi_M}$} 
\end{figure}

Stress here, this continuation is performed {\bf without} further deforming the contour  in Fig. {\ref{contr}} and considering contributions from complex
branch points observed in \cite{MAR1}. 
In other words, we know the prescription for the transformation of equation between two different spaces- Minkowski and Euclidean-,  however the timelike Minkowski solution {\bf is not} an analytical continuation of the Euclidean one.  
The numerical results on  Minkowski QED2+1 have been obtained first time in the paper 
\cite{SAUBAT2009} and here it is presented in figure \ref{jedna} and \ref{dva} Apparently, the mass function  $B$ would be non-holomorphic in the beginning $p^2=0$ if one tries to interpret the Euclidean solution as an Minkowski spacelike solution. Therefore the Wick rotation is not valid here.
In the limit $e^2>>m$ we  get chiral symmetry breaking phase
$\phi_{M_{DCSB}}=88^o$.

Until now a direct Minkowski solution of $B$ for spacelike argument is unknown to us (attempting to put similar ideas in a game, e.g. using the arguments of GFs as the integration variables,  we would certainly factorize some singularities that remain isolated in different subregions of Minkowski space). However to deal with this problem, one can employ  correct analytical continuation of the obtained timelike solution (so the strategy is completely opposite to the standard "text book" procedure, here we get the timelike solution at first and then we can construct the solution at spacelike region).
In given approximation the analytical continuation is a very straightforward if one takes $ln\rightarrow Ln$. The resulting equations reads
\be \label{cont}
B(-p^2)=m+\frac{ie^2}{4\pi^2\sqrt{p^2}}\int\limits_0^{\infty} dk^2 S_s(k)
arcsin\frac{2kp}{(k^2+p^2)} 
\ee
where $S_s$ in the rhs. of Eq. (\ref{cont})  is defined in timelike region. The numerical integration should also be straightforward and will be done elsewhere.

The second very important observation is that we get no real pole in the electron propagator.
Actually  $B$ is complex for all considered (real) $p^2$ and if bare mass $m$ is not  very large it has relatively large non-zero imaginary part preventing thus the presence of a real pole. No real pole implies no free particle solution,  which is the simple way how confinement of QED2+1 electron 
is reflected by behaviour of the GFs.

\section{DCSB in large $N_f$  QCD}

Up to date there is no known direct nonperturbative Minkowski space study of a pure Yang-Mills theory, e.g.  $N_c=3$ pure gluodynamics. There are attempts in the literature that at least questioned the analytical structure
of GFs in QCD. Serious study  of analytical continuation of the Euclidean  quark propagator performed in 
\cite{MAHO1992} shows  branch points in the complex plane of $k^2$. The branch points are located in the interior and as well as in the exterior of Wick rotation contour, however the  positions and number of  singularities found depends on the details of the interaction. Recall  also that the character of  singularities is unknown in  general, very likely they are not simple isolated poles. The truncation of SDEs is a necessary approximation in  any case and discussion of the effect of a truncation is plausible, however it seems quite natural that a strong enough interaction  likely leads to the absence of a real branch point in the quark propagator. Also the naive numerical fits and indirect estimates  based on the behaviour of Schwinger functions made for instance  in \cite{FADM2004}  signaled that the quark propagator has a more complicated analytical structure. Note also that that the fit used in  \cite{FADM2004} is in  contradiction with the  Wick rotation. As in the previous case I am going to discuss the fermion SDE, but now in 3+1 Minkowski space. Using a certain simplification the model has a confining phase characterized by absence of a real fermion  propagator pole. In this respect we are confirming  proposals by Fukuda, Kugo \cite{Fukuda}, but reaching  quantitatively very different result since avoiding using Euclidean space calculation at intermediate step.

In order to get some first solution we consider large $N_f$ QCD. Here the number of massless quarks is  larger then in the nature, decreasing considerably low $q^2$ running, but limited from the up in order to preserve asymptotic freedom. Here we assume that Euclidean theory is at least a good guide for the estimate of  the magnitude of an  effective running coupling. So for instance $N_f=7,8$ is a reasonable estimate for $N_c=3$. For this number we still should get chiral symmetry breaking. Typically in this and similar models the form factor and related effective coupling do not run but mildly change in the low scale (defined by the low energy mass of the quarks $m<<\Lambda$) and vanishes at high scale in accordance with asymptotic freedom. We basically use this scenario here we consider  the same model as in the paper \cite{KUSH2006}. It is also notable, these Technicolor (TC) like model remains viable scenario for electroweak symmetry breaking without inclusion of the scalar Higgs field \cite{KUSHYA2007}.

In our Minkowski study we model the effective running charge as follows      
 \bea 
&&{\mbox{for}}\,\, |q^2|<\Lambda^2 : \alpha_{TC}(q^2)=\alpha^* \, \,;
\nn \\
&&{\mbox{for}}\,\, |q^2|>\Lambda^2 \alpha_{TC}(q^2)=0 \,\, ,
\nn
\eea
where constant $\alpha^*$ is large enough to get DCSB. This coupling enters the ladder approximated SDE 
which in Landau gauge and $A=1$ approximation reads
\be \label{dement}
M(p^2)=m+i C\int\frac{d^4k}{(2\pi)^4}S_s(k)\frac{\alpha_{TC}(q^2)}{q^2}\, \, .
\ee
The general structure of the quark propagator is given again by 
(\ref{cpxS}) and the dynamical mass function is simply $M=B$.
$C$  is the  constant stemming from  Casimir of given representation of Non Abelian gauge group
and the prefactor from the Lorentz and Dirac algebras.

 Minkowski momentum space integral which appears in our SDE is just an one more spacelike dimension extended analogue of $I$ considered for QED2+1, here it explicitly reads
\be
I[G,S;p)=i\int\frac{d^4k}{(2\pi)^4}G(k^2)S(q^2) 
\ee
where $G=\frac{\alpha_{TC}(q^2)}{q^2}$ in our case.

Lorentz invariance dictates that $B(p^2)$ is the function of only$p^2$ for any choice of  $p^{\mu}$, so to get a solution it seems to be advantageous to use some simple choice of configuration, for spacelike 
$p^{\mu}=(0,0,0,p)$  while for timelike $p^2$  $p^{\mu}=(p,0,0,0)$.

In the usual Euclidean studies  it is desired to use arguments of GFs as an integration variables.
Unfortunately  for a spacelike $p$ it leads to factorization of an awfully divergent or singular integrals 
which are completely independent on  the behaviour of  GFs inside the integrals .
Actually one can arrive to  the following singular integral
    
\bea
&&for \, \,p^2<0,
\nn \\
&&I[G,S;p)\simeq \int_{-1}^{1}d z \left(\int_0^{\infty} dk^2\int\limits_{k^2+p^2}^{\infty sgn (z)}+
\int^0_{-\infty} dk^2\int\limits_{k^2+p^2+2\sqrt{-p^2}\sqrt{-k^2}z}^{\infty sgn (z)}\right)
\nn \\
&&\frac{1}{z^3}\frac{1}{k^2+a/z^2}\, K\, G(k^2) S(q^2)\nn
\\ 
&&K=\frac{(q^2-k^2-p^2)^2}{8(\sqrt{-p^2})^3 }\, \, ; \, \,
a=\frac{(q^2-k^2-p^2)^2}{2(\sqrt{-p^2}^2)}
\nn
\eea
so the problem is very badly defined. It is almost redundant to say that in perturbation theory this is just the Wick rotation with  its $\ep$ prescription which regularize unwanted singularities.     
We do not know a different Minkowski space regularization which preserves all required symmetry (Lorentz and gauge if required) for spacelike $p^2$, however for positive $p^2$ one can arrive to a more optimistic formula, which when used in our SDE
(\ref{dement}) gives the following integral equation: 
\bea \label{oneloopskeleton}
M(p^2)&=& m+\frac{i C}{16 p^2 \pi^3} \int_0^{\infty}d k^2
\left[\int\limits_{(k+p)^2}^{\infty}d q^2 \sqrt{\Delta(q^2,k^2,p^2)}
+\int\limits^{(k-p)^2}_{-\infty}d q^2\sqrt{\Delta(q^2,k^2,p^2)}\right.
\nn \\
&+&\left.\int^0_{-\infty}d k^2\int^{\infty}_{-\infty}d q^2\sqrt{\Delta(q^2,-k^2,p^2)}\right]
[S_s(k^2)\frac{\alpha_{TC}(q^2)}{q^2} \, \, ;
 \\
\Delta(a,b,c)&=&a^2+b^2+c^2-2(ab+ac+bc)
\nn
\eea

Not similarly to QED2+1  the equation (\ref{oneloopskeleton}) remains coupled with unknown $S_s$ for the  spacelike arguments.
It requires some further approximations which together with the detail of the derivation of Eq.
\ref{oneloopskeleton} will be published in the shortly coming paper \cite{PREP}.

\begin{figure}
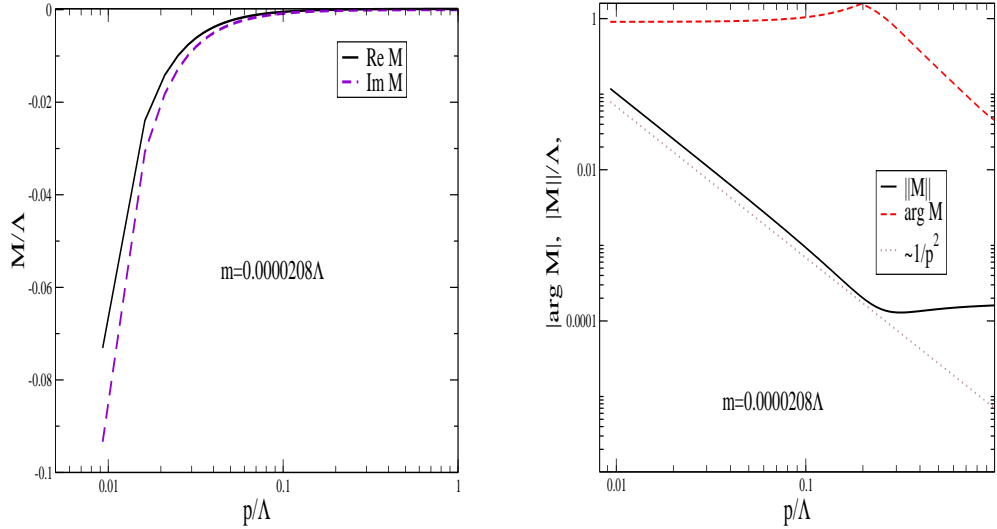

\vspace{1cm}
\epsfig{figure=Minkowskimass.eps,width=6truecm,height=7truecm,angle=0}
\hspace{1cm}\label{details}
\epsfig{figure=Minkowskimass3.eps,width=6truecm,height=7truecm,angle=0}
\vspace{0.5cm}\caption[caption]{Left panel shows real and imaginary parts of the dynamical mass separately, the right panel display the same as previous figure but in log-log scale for the both quantity, mass is rescaled by $\Lambda$, } 
\end{figure}

\begin{figure}
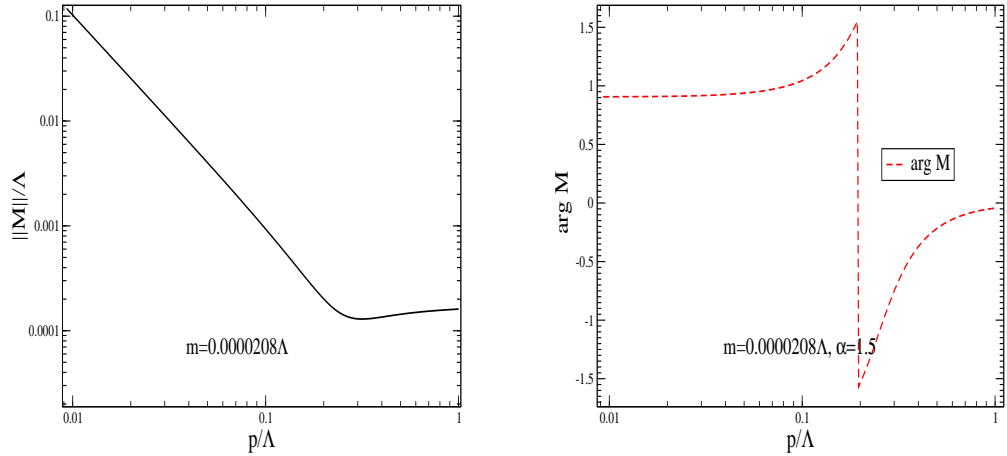

\vspace{1cm}
\epsfig{figure=Minkowskimass2.eps,width=6truecm,height=6truecm,angle=0}
\hspace{1cm}
\epsfig{figure=Minkowskimass4.eps,width=6truecm,height=6truecm,angle=0}
\vspace{0.5cm}\caption[caption]{\label{baremass} On the left panel we plotted the magnitude $\phi$ (in rad) of dynamical quark mass in 3+1 Minkowski dimension, the phase $\phi$ is on the right, note $ M=||M||e^{i\phi}$.  }
\end{figure}

The present numerical solution has been obtained for the effective coupling  $\alpha^*=1.5$
and  the constant prefactor $C$ has been absorbed into it. Until now we did not perform an ultimate study of large parameter space, but we can conclude that we have never observed 
nontrivial solution for exactly zero bare mass $m=0$.
Since the value can be arbitrarily small we were slowly decreasing $m$ during the iteration process to a desired value. The bare mass we used in our calculation is presented in the figure $\ref{baremass}$. All results are scaled with respect to $\Lambda$ which is quite natural scale (recall experimental $\Lambda_{QCD}^{2N_f}=450 MeV$ while expected $\Lambda_{Technicolor}=1-10 TeV$. 
The solution we get is completely stable and in principle  we can reach it with an arbitrary high accuracy.
The most apparent fact is the infrared behaviour, the mass simply goes to infinity as $1/p$ there, in other words constituent (techni)quark mass (even as absolute value) is not well defined quantity since $lim_{p^2\rightarrow 0}   ReM(p^2), Im M(p^2) \rightarrow \infty$. At the time being we do not know whether this is a pathological feature of the Minkowski metric in used, or it it is a physically acceptable scenario for strong coupling 3+1 dimensional theories. This remains to be a subject of future studies and confirmations (based for instance on the bound state solutions).

Further, like in QED2+1, there is no physical pole mass, the mass function is complex preventing the existence of a real pole of the propagator. Following the  LSZ-reduction formula for S-matrix, the matrix elements between quark states is zero, since the limit $lim_{p^2\rightarrow m^2}(p^2-m^2)S(p^2)=0$ for any considered real $m^2$, or in better words: free quarks do not exist.

To conclude in one sentence, there are  many issues to be clarified in future Minkowski space nonperturbative studies, but the clue to be followed already exists and the perspectives are slowly opening.

\end{document}